# TrialView: An AI-powered Visual Analytics System for Temporal Event Data in Clinical Trials

| Zuotian Li | Xiang Liu | Zelei Cheng | Yingjie Chen | Wanzhu Tu | Jing Su |
|---|---|---|---|---|---|
| Purdue University | Indiana University | Northwestern University | Purdue University | Indiana University | Indiana University |
| li4007@purdue.edu | xl95@iu.edu | zelei.cheng@northwestern.edu | victorchen@purdue.edu | wtu1@iu.edu | su1@iu.edu |

## Abstract

*Randomized controlled trials (RCT) are the gold standards for evaluating the efficacy and safety of therapeutic interventions in human subjects. In addition to the pre-specified endpoints, trial participants' experience reveals the time course of the intervention. Few analytical tools exist to summarize and visualize the individual experience of trial participants. Visual analytics allows integrative examination of temporal event patterns of patient experience, thus generating insights for better care decisions. Towards this end, we introduce TrialView, an information system that combines graph artificial intelligence (AI) and visual analytics to enhance the dissemination of trial data. TrialView offers four distinct yet interconnected views: Individual, Cohort, Progression, and Statistics, enabling an interactive exploration of individual and group-level data. The TrialView system is a general-purpose analytical tool for a broad class of clinical trials. The system is powered by graph AI, knowledge-guided clustering, explanatory modeling, and graph-based agglomeration algorithms. We demonstrate the system's effectiveness in analyzing temporal event data through a case study.*

## 1. Introduction

Randomized controlled trials (RCT) are designed experiments of human subjects for evaluation of the efficacy and safety of therapeutic interventions. Trials are designed to generate actionable, reliable, and reproducible evidence in support of specific treatment strategies, usually in comparison to standard care. RCTs are the gold standard for therapeutic evaluation and the foundation of evidence-based medicine.

RCTs generate a plethora of information, including outcomes, detailed treatment processes, participant characteristics, laboratory measures, and adverse events. Trial data are typically collected as longitudinal occurrences of events of individual participants. Aggregating and summarizing individual sequences of these individual events, however, pose great analytical challenges because of the inherent heterogeneity and the complexity of temporal patterns. Traditional statistical methods are often limited in their capacity to deal with heterogeneous high-dimensional and sequential event data.

Various analytical approaches have been developed to address these challenges, including permutation tests and trend-based analysis, agent-based simulation techniques, and artificial intelligence (AI) models such as recurrent neural networks and convolutional neural networks. However, current approaches cannot intuitively depict the event transition trajectories for care providers and clinical investigators who rarely possess knowledge of data science. To the best of our knowledge, there are no general-purpose and user-friendly tools for summarizing and visualizing the longitudinal patterns of participants' event sequences in RCTs.

Herein, we attempt to bridge this gap by developing TrialView, an interactive visualization system based on the practical needs of clinical trialists, care providers, and medical investigators. The main features of the system are designed with their input. The goal is to provide a clear and concise display of the trial events data and enable users to explore and recognize discernable patterns and communicate the findings with stakeholders.

TrialView provides a comprehensive solution that combines explainable AI techniques and visual analytics to effectively analyze and interpret event sequences in RCTs. The system was developed as part of the research infrastructure of the Alcoholic Hepatitis Network (AlcHepNet), which is a national consortium consisting of eight clinical sites and ten translational laboratories. AlcHepNet is a perfect platform for such development as it conducts multicenter RCT and observational studies. The network personnel include clinical trialists, data managers and analysts, research staff, care providers, and medical investigators, thus presenting a broad user base for developing and testing the system as their needs and expectations are often different.

This work makes three key contributions: 1) Through hierarchical task analysis, we identify a set of tasks and requirements to fulfill the requirements of caregivers and researchers on trial data analysis. 2) We develop an explainable graph AI method that incorporates baseline lab test results and temporal event sequences to cluster the cohort, resulting in a comprehensive grouping of patients based on their clinical characteristics and disease progression patterns. 3) We present a multi-view visualization system for insights into individual and group patterns in the RCT data. These intuitive representations support effective exploration, clustering, and summarization of the data. We validate the practical utility of our system through a case study. By demonstrating its effectiveness in real-world scenarios, we establish the system's value in supporting clinical researchers, care providers, and other stakeholders in their decision-making processes.

## 2. Related work

### 2.1. Visual analytics in clinical trial

Visual analytics plays a crucial role in trial management and decision-making by integrating diverse data sources to enable interactive exploration of patterns, trends, outliers, and relationships. Prior studies have emphasized the significance of effective data analysis in identifying potential adverse events related to investigational drugs (Wang et al., 2020). Visualization serves as a valuable bridge between data scientists and clinical researchers, facilitating seamless communication and knowledge exchange.

In their study, Wang et al. (2020) propose analysis approaches for general safety review and specific safety topics of interest. In contrast, our approach focuses on a comprehensive temporal summary of the data, capturing broad data patterns and temporal transitions. Lamy et al. (2017) employ dynamic tables and rainbow boxes to present comparative drug information, while ClinOmicsTrail (Schneider et al., 2019) integrates clinical and omics data using visual analysis tools such as radar plots, sunburst plots, and tables for breast cancer treatment stratification. TabuVis (Nguyen et al., 2012) offers a high-dimensional solution for visualizing metadata, subject information, and flow cytometry files through scatterplots and filtering. Lamy (2020) compares four visual analysis techniques for adverse event rates in clinical trials, namely horizontal stacked bar graphs, vertically stacked bar graphs, area proportional flower glyphs, and star glyphs, and concludes that horizontal bar graphs and flower glyphs are more effective. However, these studies do not fully address the comprehensive analysis of broad data patterns and temporal progression during treatment, which is a key aspect addressed by our approach.

### 2.2. Event sequence and progression visualization

Our clinical trial data for Alcoholic Hepatitis (AH) encompasses a wide range of events with different types, orders, and durations throughout the trial. To aid users in comprehending and uncovering patterns within this extensive dataset, visualization plays a crucial role. Guo et al. (2022) comprehensively surveyed event sequence visualization works spanning across timeline-based, Sankey-based, hierarchy-based, matrix-based, and graph-based visualizations. Moreover, Ledesma et al. (2019) have demonstrated the usability and effectiveness of health timeline visualization in understanding patient health trajectories. Inspired by this, we incorporate timeline visualization into our information system to enhance the exploration and analysis of AH clinical trial data.

Pathway visualization includes sequence alignment (Li & Homer, 2010), progression analysis, and interactive exploration. IDMVis (Zhang et al., 2019) enables users to fold and align records to extract event sequence patterns. However, it is designed for individual patient data, whereas our view extends this capability to explore patterns across entire cohorts. DPVis (Kwon et al., 2021) applies Pathway Waterfall to display state transition paths using parallel beeswax plots and trajectory lines connected by force edge bundling. EventFlow (Monroe et al., 2013) supports searching, summarizing, cohort selection, simplification, and analysis of population-level patterns, but it lacks a clear summary of the transition between events. ThreadStates (Wang et al., 2021) utilizes Sankey-based visualization, scatterplots, and glyph matrices to identify disease progression states through human-in-the-loop learning. Yet, its visualization of states is not time specific. These previous visualizations are absent of temporal aspects. In contrast, our approach offers a population-level perspective, highlighting event transition patterns.

### 2.3. Clustering the patient events pattern

Clustering similar sequences and extracting concise representations of data play a crucial role in understanding complex temporal clinical patterns from longitudinal RCT data. Participant clustering helps identify sequences that share similar progression paths. Current clustering algorithms are based on similarity and dissimilarity like k-means (Hartigan & Wong, 1979), and hierarchical clustering (Johnson, 1967). Wongsuphasawat et al. (2009) measure the similarity between patient histories based on aligning temporal categorical data. Guo et al. (2019) propose an unsupervised status analysis technique to discover semantically relevant progression status as well as characteristic events. DPVis (Kwon et al., 2021) employs hidden Markov models to capture the dynamics of disease progression. These methods show limited performance when multiple data domains such as patient demographics, baseline lab test biomarkers, and temporal events are used. Graph artificial intelligence models such as graph convolutional networks (Fang et al., 2021) and graph transformers (Tang et al., 2023) have demonstrated superior capacity over these methods when incorporating multimodality data for clustering. Explainable AI approaches such as Grad-CAM (Selvaraju et al., 2017) further gain insight into AI models.

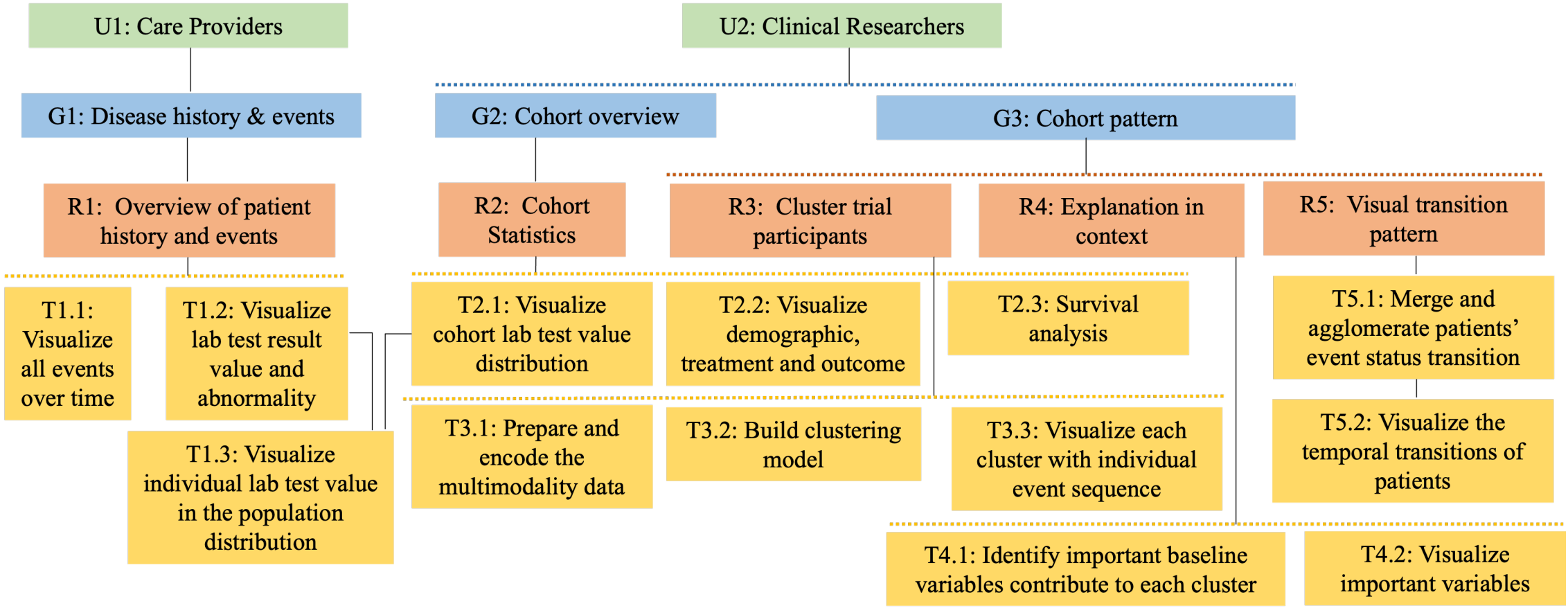

**Figure 1. Task analysis.** G: Goals; U: Users; R: Requirements; and T: Tasks.

## 3. Task analysis

Inspired by IDMVis (Zhang et al., 2019) and Salmon et al. (2020), we utilize hierarchical task analysis (HTA) to enhance our framework for specific users. HTA breaks down complex tasks into smaller sub-tasks, providing a systematic understanding of workflow steps, decision-making, and interactions. The task analysis approach aligns well to categorize and comprehend the diverse range of activities that users engage in when interacting with visualizations. HTA comprehensively examines tasks, ensuring that the information system accommodates a broad range of user requirements and objectives. We use HTA to design a user-centered system aligned with their needs, workflows, and goals. Through collaboration with care providers (U1) and clinical researchers (U2), we optimize the framework to address pain points and enhance patient care and research outcomes. As a result, we determined the following three goals:

The first goal is to present the patient's medical history and events (G1) including laboratory test results and occurrences of adverse events to care providers (U1). Such information is crucial for understanding the patient's baseline health status, pre-existing conditions that may influence their response to treatment, and the events during the trial.

Longitudinal data from clinical trials will provide clinical researchers (U2) clinically meaningful insights into disease progression patterns. The second goal is to reveal the distribution of population characteristics and outcomes (G2). Such visualization helps identify demographic factors, comorbidities, or genetic variations that may impact the efficacy of certain treatments, and better tailors interventions to specific subgroups and improve overall patient care.

The third goal is to know the course of the population progression pattern (G3). This information is essential for predicting disease trajectories, identifying potential risk factors, and developing targeted interventions. By tracking the progression pattern of the population over time, healthcare providers can gain insights into disease progression rates, treatment response, and the effectiveness of interventions. Such knowledge allows them to make informed decisions about resource allocation, preventive measures, and early interventions.

Each of these goals is associated with specific target user groups. G1 primarily targets care providers (U1) who need a comprehensive understanding of the patient's medical history to guide their treatment decisions. G2 is relevant to clinical researchers (U2) who aim to analyze population-level data to identify trends, disparities, and potential areas of improvement in healthcare delivery. G3 caters to clinical researchers (U2) who require population-level progression patterns to develop strategies for disease prevention, resource allocation, and public health interventions.

We discussed these three goals with care providers and clinical researchers and summarized five critical requirements as follows:

(R1): Provide an overview of the individual's clinical history, laboratory test results, and events.

(R2): Summarize cohort statistics. Researchers want to know the distribution of demographic and lab test values and compare the two treatment outcomes.

(R3): Cluster the common group. Aggregate the patient history and intend to find trajectory clusters and compare the patient experience clusters.

(R4): Provide interpretation in the context of treatment, demographics, and lab values. Once the patients have been clustered by disease progression, researchers would wonder about the relationship between lab test results, demographics and specific trajectories. The system should provide an explanation to show the contribution of important baseline biomarkers to the trajectory.

(R5): Summarize each cluster as a state transition pattern. Each of the states should have a clinical meaning, and the visualization should show the order, the time of the state transition, and the number of patients in that transition.

Based on the identified goal and requirements, we decompose the tasks as shown in Figure 1.

# 4. Approach

## 4.1. Data processing

The data (Table 1) used in this work is composed of baseline features and follow-up events of AH patients ($N = 147$) who participated in the RCT (ClincialTrials.gov Identifier: NCT04072822) carried out by AlcHepNet. Patients were randomly assigned to two treatment arms for 90 days, with 73 and 74 patients receiving Treatment A and Treatment B, respectively. Demographic, vital, behavioral features, and laboratory tests were obtained at baseline ($t = 0$). Patients were followed up for 180 days for events including death, liver transplantation off study, early stop of treatment, as well as the episodes of acute kidney injury (AKI, defined in a general sense, including kidney injuries that persisted up to 180 days), infection, and other adverse events (OAE).

**Table 1. Data elements used in this study.**

| Category | Number of features | Features |
|---|---|---|
| Demographics | 3 | Sex, Race, Age |
| Vitals | 1 | Body mass index (BMI) |
| Lab tests | 15 | Meld Score (MELD), Hemoglobin (Hb), White Blood Cells (WBC), Platelets, Mean Corpuscular Volume (MCV), International Normalized Ratio (INR), Prothrombin Time (PT), Albumin (Alb), Total Bilirubin (T Bil), Direct Bilirubin (D Bil), Creatinine (Cr), Alanine transaminase (ALT), Aspartate aminotransferase (AST), Alkaline Phosphatase (ALK), Total Protein (TP) |
| Drinking behaviors | 2 | # of drinks in the past 30 days, # of drinking days in the past 30 days. |
| Treatment | 2 | Treatment A and Treatment B |
| Events | 7 | Treatment, Other Adverse Event (OAE), Infection, Acute kidney injury (AKI), Off Study, Liver Transplant, Death |

Since multiple events may occur at the same time, clinical events are further merged into 9 mutually exclusive event statuses denoted as $E_k$ and $k = 1, \dots, 9$ with the following cascade trump rule: Liver transplant/Death > Off study > AKI + Infection > AKI > Infection > OAE > Treatment + OAE > Treatment > No event. For example, on a specific day, if a patient suffers from both infection and OAE, the summarized event status will be "Infection".

The definition of event sequence and its mathematical notation are based on the 9 summarized event status. For a patient $i \in \{1, \cdots N\}$, the baseline features $r$ are denoted as $b_{i,r}$. The event sequence of this patient is represented as a sequence $\boldsymbol{\tau_i} = \{s_{i,t}\}$, where $t = 0,1, \cdots, 180$ denotes a specific follow-up day, and $s_{i,t}$ represents one of the 9 event statuses.

## 4.2. Clustering analysis

Patients are clustered into subgroups according to their baseline features and trajectories and cluster-specific commonalities and patterns are visualized. Our system provides two options: data-driven clustering using graph artificial intelligence and knowledge-guided clustering.

**4.2.1. Data-driven clustering with graph AI.** A graph transformer autoencoder (Figure 2) is used to cluster patients according to the similarity of both baseline features and status sequences. The original data is first represented as a patient similarity graph $G = (V, E)$. The graph is constructed according to the baseline features, with each node $v_i \in V$ represents a patient, and an edge $e_{i,j}$ represents two patients with similar baseline features, i.e., $\left\|\{b_{i,r}\}, \{b_{j,r}\}\right\| \le \sigma$. The event status sequence $\boldsymbol{\tau_i}$ is used as the node property. Then a latent representation of patients' baseline and follow-up data is learned. Briefly, for a patient $v_i$, the propagation of the graph transformer from the $l$ layer to the $l+1$ layer is defined as:

$$\boldsymbol{h}_i^{(l+1)} = \mathrm{ReLU}\left(W_i^{(l)} \boldsymbol{h}_i^{(l)} + \sum\nolimits_{v_j \in \mathcal{N}(v_i)} \alpha_{i,j} V_j^{(l)}\right)$$

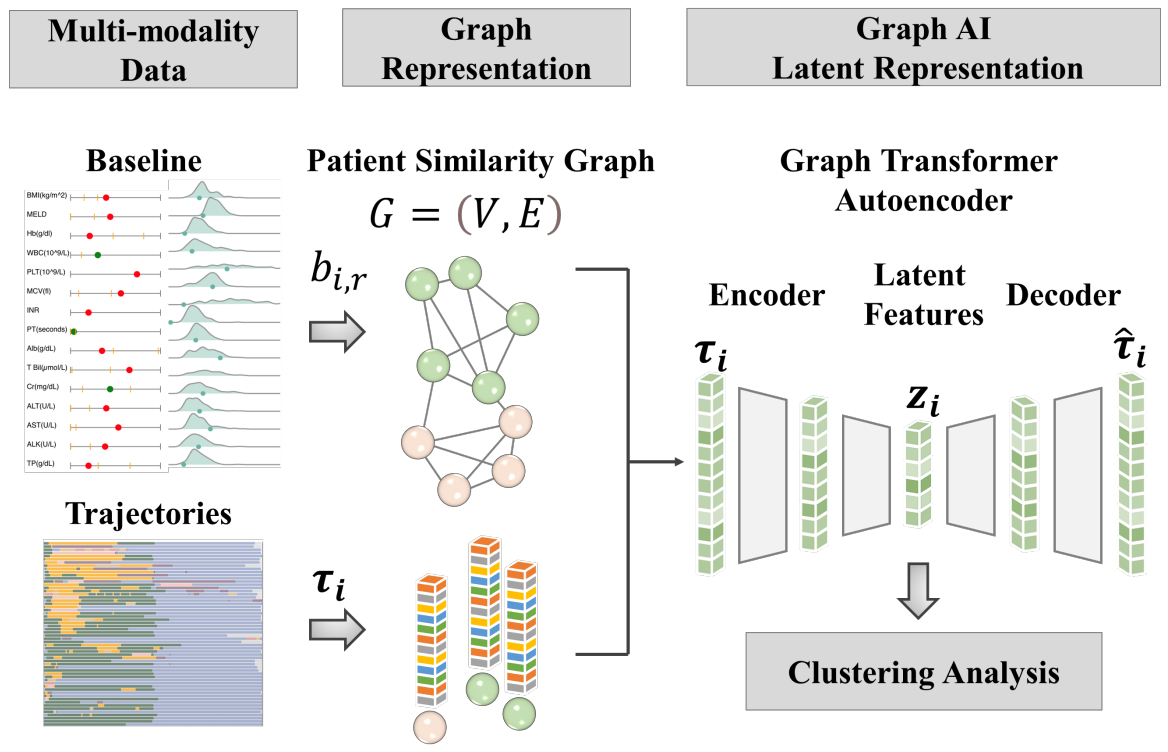

**Figure 2. Graph transformer clustering.**

, where the rectified linear unit (ReLU) is used as the nonlinear gated activation function, $\mathcal{N}(v_i)$ represents the neighbor nodes of $v_i$, and $\boldsymbol{h}_i^{(0)} = \boldsymbol{\tau_i}$. The attention module is defined as:

$$\alpha_{i,j} = \mathrm{softmax}\left(\frac{\langle Q_i^{(l)}, K_j^{(l)}\rangle}{\sum_{u \in \mathcal{N}(i)} \langle Q_i^{(l)}, K_u^{(l)}\rangle}\right)$$

, where:

$$\text{query: } Q_i^{(l)} = W_Q^{(l)} \boldsymbol{h}_i^{(l)} + b_Q^{(l)}$$
$$\text{key: } K_j^{(l)} = W_K^{(l)} \boldsymbol{h}_j^{(l)} + b_K^{(l)}$$
$$\text{value: } V_j^{(l)} = W_V^{(l)} \boldsymbol{h}_j^{(l)} + b_V^{(l)}$$

and $\langle Q, K\rangle \equiv \exp\left(Q^T K \big/ \sqrt{\dim\left(\boldsymbol{h}_i^{(l)}\right)}\right)$.

Hyperparameters are determined by finetuning. The patient similarity graph was constructed as a k-nearest neighbor network ($k = 10$). The dimensions of the decoder layers are 78 and 36, and the encoder is symmetric to the encoder. The data is split into the training, test, and validation subsets at an 8:1:1 ratio. The model is trained using the mean squared error loss, with the batch size of 512, 300 epochs, and the Adam optimizer at a learning rate of $1e-5$.

**4.2.2. Knowledge-guided clustering.** Patient event statuses are encoded based on clinical domain knowledge obtained through interviews with clinical researchers. Specifically, we categorized the four major endpoints of the clinical trial into distinct groups: death or liver transplantation (coded as 20), off-study (coded as 15), adverse events (coded as 2 – 5), and no event (coded as -5). Subsequently, patients were clustered using weighted Ward's agglomerative hierarchical clustering, chosen for its robustness with coded data (Murtagh & Legendre, 2011).

## 4.3. Learning contributing baseline features with a graph Grad-CAM model

To reveal baseline features that contribute to each identified cluster, we use a graph gradient weighted class activation map (graph Grad-CAM) model on the graph transformer autoencoder, as its performance has been proved in various applications (Selvaraju et al., 2017). Briefly, a 2-layer multi-layer perceptron (MLP) neural network is used to predict the patient's cluster membership, with the latent representation learned from the graph autoencoder as input, the identified clusters as ground truth, and the cross-entropy loss function. For each identified cluster $c$, the predicted possibility of patient $i$ belonging to this cluster is $y_{i,c}$. The importance of baseline feature $b_{i,r}$ for this patient is:

$$\alpha_{i,r}^c = \mathrm{ReLU}\left(\frac{\partial y_{i,c}}{\partial b_{i,r}} \cdot b_{i,r}\right)$$

, and the overall importance of baseline feature $b_{i,r}$ for cluster $c$ is:

$$\alpha_{i,r}^c = 2 \cdot \mathrm{softmax}\left(\frac{\alpha_{i,r}^c}{N}\right) - 1$$

. The patient-level importance scores delineate the contribution of each baseline feature of each patient for his or her cluster membership. The cluster-specific importance scores describe the cluster-level important baseline features. We employed these importance scores to create a heatmap, as depicted in Section 5.2, to assist the user in identifying crucial features.

## 4.4. Status agglomeration

Having established the representation of individual trajectories as mentioned earlier, our focus now shifts to examining event transition patterns from a population perspective over 180 days. For a specific cluster $c$ and the $n$ patients belonging to this cluster, the corresponding event sequences are $\boldsymbol{\tau_1}, \cdots, \boldsymbol{\tau_n}$ . An event status agglomeration algorithm is developed to reveal the transition patterns.

Firstly, the status transition of an event status across the time series can be represented with a one-directional chain model, with each node denoting a specific time block, and a directional edge linking two neighbor time slots. As shown in Figure 3, we first construct a $9 \times 180$ status matrix, with each row representing a distinct event status $E_k$ and each column denotes a specific day $t$ in the duration of the study period. The transitions during the whole trial can be denoted as a chain of 180 time blocks (i.e., nodes). Specifically, for the day $t$ and a given event status $E_k$, the relevance of this day to $E_k$, the transition statuses from the previous day and to the next day can be represented as a triplet ($\mathrm{Num}_t$, $\mathrm{In}_t$, $\mathrm{Out}_t$) that encapsulates the state transition information, where $\mathrm{Num}_t$ is the number of patients with event status $E_k$ at time $t$, $\mathrm{In}_t$ is a $9 \times 1$ vector representing the status transition from the previous time block to the current one, and $\mathrm{Out}_t$ is the transition vector to the next time block. For example, the transition from the event status $E_m$ at time block $t-1$ to $E_k$ at time block $t$ is represented as $\mathrm{In}_t[m] = \sum_i^n (s_{i,t-1} = E_m\ ,\ s_{i,t} = E_k)$. The flow of patients into and out of the event status $E_k$ at time block $t$ remain balanced, that is, $\sum_m \mathrm{In}_t[m] = \sum_m \mathrm{Out}_t\,[m] = \mathrm{Num}_t$.

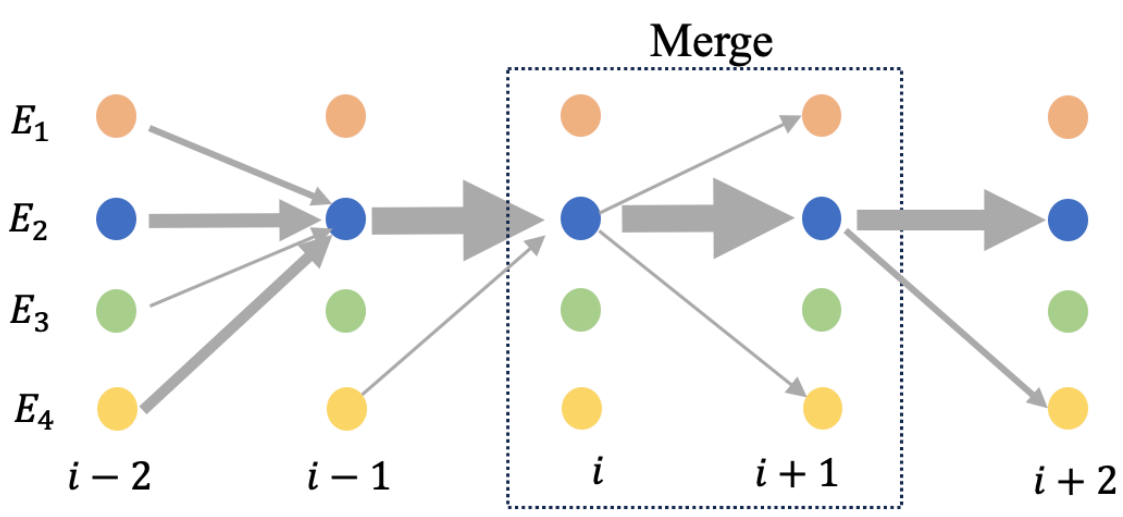

**Figure 3. Agglomeration of neighbor time blocks.**

Then the chain model can be simplified by agglomerating neighbor time blocks that show similar transition patterns. We develop an agglomeration algorithm to merge time blocks with similar state transition patterns from a cohort view. Algorithm 1 describes the merging process. The overall transition pattern at a time slot $t$ is represented by a $19 \times 1$ vector $Q_t|_{E_k} \equiv [\mathrm{Num}_t,\ \mathrm{In}_t,\ \mathrm{Out}_t]^T$ The similarity of transition patterns between two neighbor time slots $t$ and $m$ for an event status $E_k$ can be measured by

**Algorithm 1** MERGE_LINKEDLIST

**Input:** A long chain of LinkedList $\mathcal{L}$
**Output:** A short chain of LinkedList $\mathcal{L}$
**Initialization:** $node \leftarrow \mathcal{L}.head$
**while** $\mathcal{L}.len > 1$ **do**
  **while** $node.next$ exists **do**
    $node.sim \leftarrow \mathbf{J}(node['vec'], node.next['vec'])$
  **end while**
  $node \leftarrow node$ of $\max(node.sim)$
  **if** $node.sim > \delta$ **then**
    merge $node$ and $node.next$
  **else**
    break
  **end if**
**end while**
**return** $\mathcal{L}$

**Algorithm 2** STAGE_AGGLOMERATION

**Input:** Event Sequences $\tau_1, \tau_2, ..., \tau_n$
**Output:** A aggregated stages $\mathcal{S}$
**Initialization:** $\mathcal{S} \leftarrow \varnothing$
**for** $Event\ E_s \in E$ **do**
  **Construct a chain of LinkedList** $\mathcal{L}$
  **for** $t \in 0 : 180$ **do**
    $\mathcal{L}.node[t] \leftarrow \{\sum_{i=1}^n (\tau_{i,t} = E_s), \mathtt{flow}(\tau_i[t])\}$
  **end for**
  add MERGE_LINKEDLIST($\mathcal{L}$) to $\mathcal{S}$
**end for**
**return** $\mathcal{S}$

Jaccard similarity $J\left(Q_t|_{E_k}, Q_m|_{E_k}\right)$. The Jaccard similarity of each time block to the next time block is first calculated. Then, similar neighbor nodes, defined as that similarity exceeds a threshold $\delta$, are merged. This process is repeated till no more time blocks are similar enough. This process is performed for every event status, as described in Algorithm 2. By merging states with similar transit patterns using a single-linked list data structure and employing the Jaccard Similarity metric, the algorithm effectively condenses the information and facilitates the analysis of common patterns within the population.

Finally, the transitions across different event statuses are determined by matching the corresponding transition patterns. Briefly, the strength of a transition from an agglomerated time slot $node_t|_{E_k}$ to a neighbor time slot $node_m|_{E_p}$ is measured by Jaccard similarity defined as: $J\left(node_t|_{E_k}, node_m|_{E_p}\right) \equiv \frac{\mathrm{Out}_t[p]+\mathrm{In}_m[k]}{\mathrm{Num}_t+\mathrm{Num}_m}$ . If the Jaccard similarity exceeds a threshold $\sigma$, a transition from is $E_k$ to $E_p$ determined.

The final transition trajectory is visualized as the transitions between agglomerated time slots, which are described in detail in section 5.3.

## 5. Visual analytics system development

We develop an interactive visual analytics system of TrialView after processing and clustering data with the above algorithms. The system's backend is developed using Python, Flask, and PyTorch, while the frontend is built on React and D3.js. The architecture and workflow are depicted in Figure 4, illustrating the components involved in the information system's backend. These components include data processing, AI/ML models, explanation, and agglomeration. The React-based frontend interface and the backend components communicate through a RESTful API interface. Figure 5 showcases the prototype interface of our system, providing a visual representation of its design and functionality.

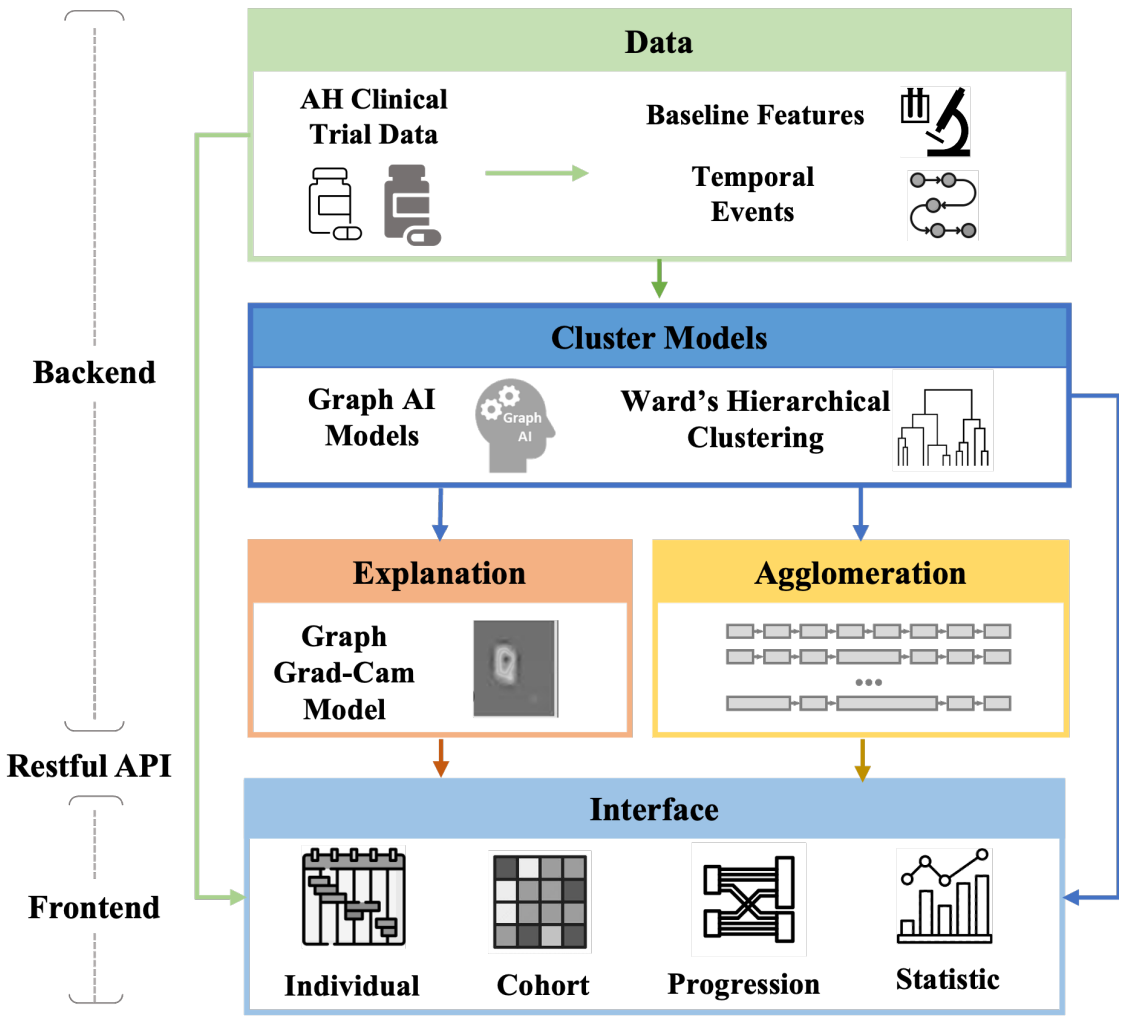

**Figure 4. System Architecture.**

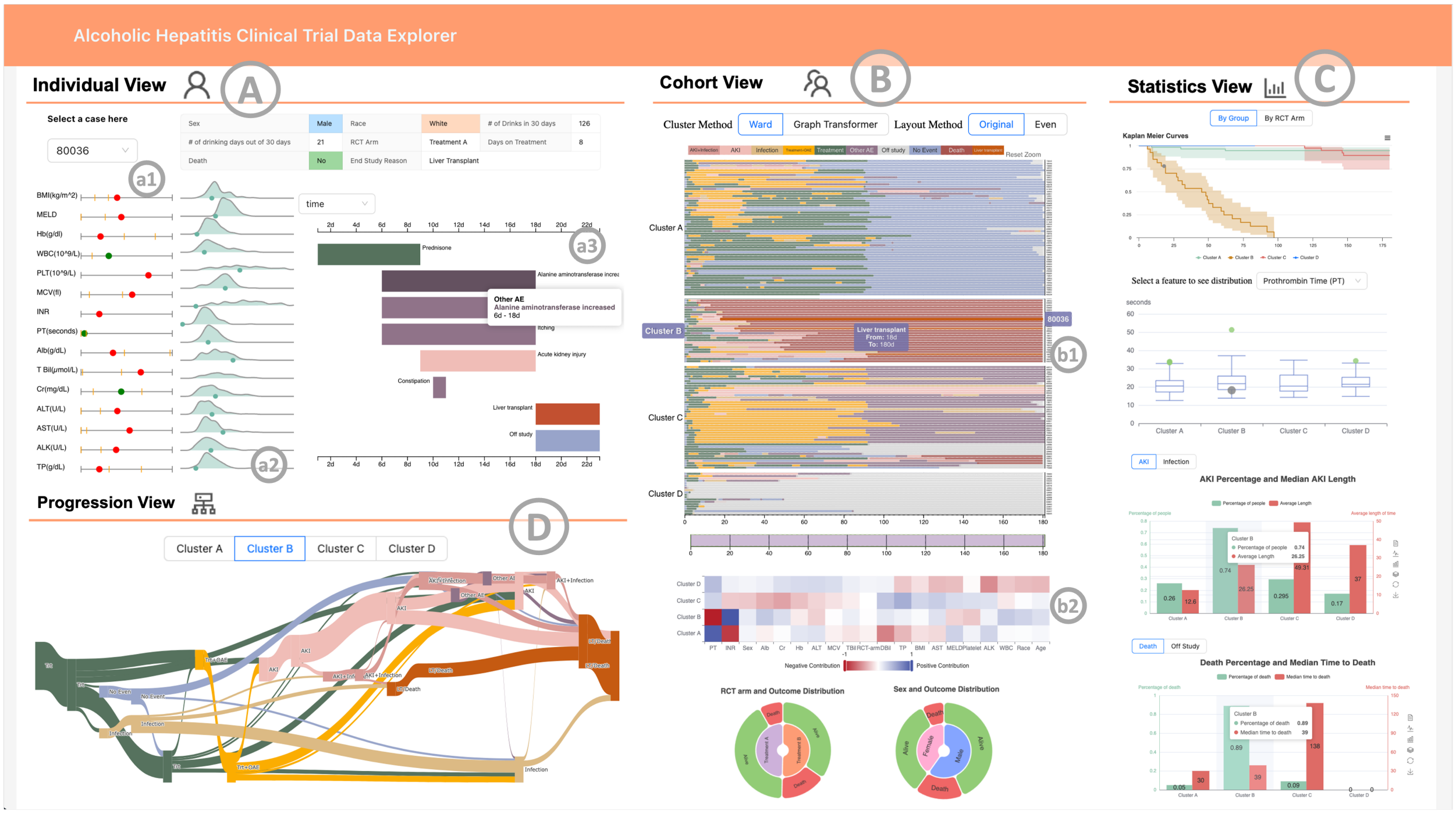

**Figure 5. System Interface.**

Expanding upon hierarchical task analysis to identify information design requirements, the four views are tailored to support these tasks. Our interface design adheres to Gestalt principles (Graham, 2008), facilitating pattern recognition for users. Additionally, colors are selected based on cultural meanings, as informed by medical experts; for example, we use red to denote death. Furthermore, we use preattentive visual perception techniques to enhance usability.

## 5.1. Individual view

In the individual view (Figure 5 A), when the user selects a participant from the queryable drop-down, the patient's demographic and pathological information is presented in the adjacent table. Figure 5(a1) highlights abnormalities in the lab tests, denoted by red dots for abnormal results and green dots for values within normal ranges. When the user hovers the mouse over a line, a context tooltip further displays the minimum and maximum values, the normal range, and the patient's specific value. Figure 5(a2) presents a ridge plot that effectively visualizes the patient's lab values in the context of the distribution within the study cohort. Additionally, the Timeline graph (Figure 5(a3)) provides a chronological overview of the patient's events, allowing the user to access information about an individual patient's baseline lab test values and track outcomes and adverse events throughout the timeline.

## 5.2. Cohort view

The cohort view (Figure 5 B) provides a comprehensive perspective on the study cohort. The timeline chart (Figure 5(b1)) displays the event trajectories of the cohort with color coding and organizes the cohort into clusters. Users can select between two clustering methods, namely Ward Hierarchy Clustering and Graph Transformer. Considering that most events occur within the initial 90 days, potentially leading to overcrowding of visual symbols in the left region, we offer a rescaled visualization so that events are evenly distributed.

To gain insight into the contributions of baseline features for different clusters (which is explained in Section 4.3), an importance heatmap (Figure 5(b2)) is used to visualize results from the explainable AI. For example, the high PT values, MELD score, and platelet counts contribute most to patients in Cluster B. Additionally, two sunburst plots depict the distribution of treatments and sex in correlation with the outcome of death or alive, located at the bottom of the cohort view. These elements collectively offer a comprehensive visualization of the cohort, enabling users to explore the clustering, event distribution, and correlations between treatments, sex, and outcomes.

### 5.3. Statistics view

The statistics view, as depicted in Figure 5(C), provides users with a comprehensive statistical overview of the cluster under consideration. It includes the survival curves with confidence intervals using the Kaplan-Meier estimator, visually representing the probability of survival at a specific time for each cluster or RCT arm. This allows users to assess the mortality risk of different clusters or treatment groups. In this view, the survival event of the selected patient in Figure 5(A) is highlighted as a grey point, providing a direct reference to the individual's survival status. Moreover, a box plot is used to portray the baseline characteristics of each cluster, offering insights into their respective contributions. The individual value associated with the selected patient is highlighted as a grey point within the box plot, emphasizing its specific placement within the distribution. To further enhance understanding, bar charts are included to present the percentage of individuals within the cluster who experienced adverse events such as AKI or infection, along with the median duration of these events. Another bar chart displays the percentage of individuals within each cluster who either died or dropped off, accompanied by the median time.

These visualizations within the statistics view effectively summarize key statistical measures, highlight individual data points, and provide comprehensive insights into survival outcomes, baseline characteristics, adverse events, and patient outcomes within the cluster under analysis.

### 5.4. Cohort progression view

The progression view, showcased in Figure 5(D), provides a concise summary of the transition patterns observed within each cluster. The plot is horizontally laid out according to the timeline, with the thickness of each line indicating the number of patients associated with that specific transition. This visualization effectively captures the collective movement and progression of patients throughout their respective timelines within the cluster, offering valuable insights into the overall pattern of transitions.

### 5.5. Interaction

To enable users to explore the data from various levels of detail and perspectives, TrialView incorporates interactive techniques that enhance the user experience. When a user selects a specific patient ID in the individual view, the corresponding individual is highlighted in both the cohort and statistics views, allowing for a seamless connection between different aspects of the data. Similarly, when the user chooses one of the two clustering methods in the cohort view, the statistics and progression views automatically update to reflect the new clusters, providing a synchronized representation of the data. To optimize the use of screen space, the system includes toggle buttons that allow users to switch between different display options. For example, in the statistics view, users can toggle between cluster and RCT arms to visualize KM curves. Common visualization interactions such as highlighting, sorting, dragging, and tooltips are supported throughout the system, enabling users to interact with the data intuitively. The interactive features enhance the usability and flexibility of TrialView.

## 6. Use case

To assess TrialView's effectiveness, we opted for a case study approach (Kitchenham et al., 1995) and engaged a hepatology researcher (H1) to employ TrialView in the analysis of data from a study focused on a novel treatment for AH. H1 aimed to gain insights at both individual and cohort levels. H1 selected participant 80036 in the Individual View (Figure 5(A)), where H1 found demographic details and a history of heavy drinking. H1 observed that several baseline lab test values were abnormal, indicated by the presence of red dots, signifying deviations from the normal range. The abnormal ALT and AST test results were also significantly high within the study cohort. Using the timeline visualization AH, the researcher explored the patient's events and discovered the ALT increase event starting at D6, followed by the AKI incidence at D9, and the liver transplantation at D18, which indicated the treatment failed. The researcher gained insights into the relations between baseline abnormal lab values, adverse events, and outcomes (*R1* in Section 3). These findings enable the researcher to better understand the new treatment's efficacy.

With a curiosity about the overall event patterns in the study cohort, researcher H1 transitions to the Cohort View. Opting for the Ward Clustering method (Figure 5(B)), H1 gained a comprehensive understanding of different event dynamics timeline charts (Figure 5(b1)): Cluster A was characterized by “No events” after 90 days and thus represented successful treatment; Cluster B indicated treatment failure as the main outcome was “Death/Liver transplant"; Cluster C were those survived with sustaining adverse events; and Cluster D reflected those who discontinued the study early (*R3*). To gain deeper insights into Patient 80036 and Cluster B, the researcher turned to the importance heatmap, cross-examined the contributing baseline features with the baseline lab values for Patient 80036, and concluded that the abnormal ALT might contribute to the treatment failure. The heatmap helps the researcher to discern the relative importance of different factors and their impact on the study population *(R4)*.

Attended by the intriguing patterns observed within a specific cluster from the cohort view, researcher H1 decided to examine the progression of the illness using the Progression View. H1 identified the trajectory of patient 80036 (Treatment to early OAE to early onset of AKI to Death/Liver Transplantation), in the context of other trajectories such as infections, late onset of AKI, etc. The researcher gained insights into the event dynamics of this patient and the cluster *(R5)*.

Subsequently, H1 navigated to the Statistics View, where critical information about patient survival is presented. Within the overall survival analysis, H1 quickly identifies that Patient 80036's survival status is censored at day 18 in Cluster B, as denoted by the highlighted grey dot (Figure 5(C)), indicating a relatively early failure of treatment among patients in Cluster B. Furthermore, the researcher leverages the box plot of lab test PT and the corresponding info in Figure 5 a1, a2, and b2 to gain deeper insights into the role of PT for the specific patient. H1 concluded that PT was less likely related to the treatment failure of this patient. Additionally, H1 appreciated that 74% of Cluster B patients experience AKI, with an average duration of 26.25 days, and 89% of Cluster B patients face mortality, with a median time to death of 39 days. Patient 80036's case was typical in Cluster B *(R2)*.

By interactively utilizing the multiple views and models offered by the TrialView system, the researcher can thoroughly explore, analyze, and derive valuable insights from the AH clinical trial data. The system empowers the researcher to make informed decisions, identify trends, and generate hypotheses for further research in the AH treatment.

# 7. Conclusion

TrialView is a comprehensive visual analysis system designed to explore, cluster, and summarize the features and temporal events observed in RCTs. Although the system was developed as a research tool for AlcHepNet, the system's functionality covers the needs of a diverse range of RCTs. This system offers four distinct views, enabling different types of users to dynamically explore the individual and cohort-level data using two clustering models. Additionally, an explanation model provides further insights into the contributions of various features to each cluster, enhancing the flexibility and interpretability of the results. The TrialView system is empowered by explainable graph AI models and graph-based agglomeration algorithms. The TrialView can also be adapted for other domains that involve temporal event sequence data, such as healthcare and business intelligence. By leveraging its flexible framework, researchers and practitioners in these fields can extract maximal benefits from the powerful analysis and visualization capabilities it offers.

For future extensions, we aim to enhance the system by supporting user-defined cluster models. This will allow users to leverage the cluster analysis and presentation functionalities and tailor the system to their specific research or analytical needs. Overall, the TrialView system represents a practically useful tool for analyzing temporal event data, with potential applications in multiple domains, and future improvements will focus on empowering users with more control over the clustering models.

# 8. Acknowledgement.

Research reported in this publication was supported by the National Institute on Alcohol Abuse and Alcoholism (U24AA026969) and the National Library of Medicine (R01LM013771) of the National Institutes of Health. We also thank Qing Tang, Carla Kettler, and Ronny Ovando at Indiana University for data preparation.